\documentclass[aps,twocolumn,pra,tightenlines,floatfix,showpacs]{revtex4}
\usepackage{amsmath}
\usepackage{amssymb}
\usepackage{float}
\usepackage{bm,times}
\usepackage{graphicx}

\DeclareGraphicsExtensions{.eps,.pdf}

\newcommand{\be}{\begin{equation}}
\newcommand{\ee}{\end{equation}}
\newcommand{\bea}{\begin{eqnarray}}
\newcommand{\eea}{\end{eqnarray}}
\newcommand{\ek}{\epsilon_{\mathbf{k}}}
\newcommand{\xik}{\xi_{\mathbf{k}}}

\newcommand{\Ek}{E_{\mathbf{k}}}

\newcommand{\phik}{\varphi_{\mathbf{k}}}

\newcommand{\sumk}{\sum_{\mathbf{k}}}

\newcommand{\sumq}{\sum_{\mathbf{q}}}

\newcommand{\Omegaq}{\Omega_{\mathbf{q}}}

\newcommand{\vk}{v_{\mathbf{k}}}
\renewcommand{\d}{\mathrm{d}}

\newcommand{\p}{\partial}
\newcommand{\D}{\mathrm{D}}
\newcommand{\Tr}{\mathrm{Tr}}

\begin{document}

\title{Stability conditions and phase diagrams for two component Fermi
gases with population imbalance}

\author{Qijin Chen, Yan He, Chih-Chun Chien, and K. Levin} 

\affiliation{James Franck Institute and Department of Physics,
 University of Chicago, Chicago, Illinois 60637, USA}

\date{\today} 

\begin{abstract}
  Superfluidity in atomic Fermi gases with population imbalance has
  recently become an exciting research focus.  There is considerable
  disagreement in the literature about the appropriate stability
  conditions for states in the phase diagram throughout the BCS to
  Bose-Einstein condensation (BEC) crossover.  Here we discuss these
  stability conditions for homogeneous polarized superfluid phases, and
  compare with recent alternative proposals.  The requirement of a
  positive second order partial derivative of the thermodynamic
  potential with respect to the fermionic excitation gap $\Delta$ (at
  fixed chemical potentials) is demonstrated to be equivalent to the
  positive definiteness of the particle number susceptibility matrix.
  In addition, we show the positivity of the effective pair mass
  constitutes another nontrivial stability condition. These conditions
  determine the stability of the system towards phase separation of one
  form or another. We also study systematically the effects of finite
  temperature and the related pseudogap on the phase diagrams defined by
  our stability conditions.
\end{abstract}

\pacs{03.75.Hh, 03.75.Ss, 74.20.-z \hfill \textsf{\textbf{cond-mat/0608454}}}

\maketitle

\section{Introduction}

Superfluidity in atomic Fermi gases has become an important research
arena
\cite{Jin3,Grimm,Jin4,Ketterle3,KetterleV,Thomas2,Grimm3,ThermoScience}
for both the condensed matter and atomic, molecular and optical physics
communities.  In these ultracold gases, via a Feshbach resonance, one
can tune the pairing interaction strength continuously from very weak in
the BCS limit to very strong in the Bose-Einstein condensation (BEC)
limit. Adding greatly to the excitement has been a recent emphasis on
experimental studies of superfluidity \cite{ZSSK06,PLKLH06,ZSSK206} in
the presence of a population imbalance between the two fermion spin
components.  This system has important consequences for other subfields
of physics including nuclear physics and high density QCD
\cite{Wilczek,LW03,FGLW05}.  From a theoretical standpoint this problem
is particularly rich and at the same time complex
\cite{SR06,Duan,Mueller,HS06,Kinnunen,Machida2,Tsinghua,PWY05,PS05a}.
Simple mean-field calculations \cite{Chien06} show that, unlike in the
equal spin mixture case, a homogeneous superfluid state is not always a
stable ground state at zero temperature ($T$).  There are a number of
different ground states to consider.  Moreover, there has been
considerable controversy in the literature
\cite{Pao,Radzihovsky,Chien06,Tsinghua} about the precise nature of the
stability conditions, associated with one mean-field ground state or
another.

It, thus, becomes particularly important to characterize and study the
stability conditions associated with polarized superfluids.  Moreover,
since there is a natural extension
\cite{ChenPRL98,ourreview,ReviewJLTP,Chien06,ChienRapid} of the standard
ground states to finite temperatures $T$, in this paper, we derive these
stability conditions for a homogeneous interacting Fermi gas superfluid
with population imbalance throughout the entire BCS-BEC crossover, and
at arbitrary $T$. We will focus our attention on a superfluid composed
of a condensate of pairs which has a zero total center-of-mass
momentum. Therefore, we will not discuss the
Fulde-Ferrell-Larkin-Ovchinnikov (FFLO) state \cite{FFLO}, which allows
condensation of pairs at finite momenta. It appears that the FFLO state
with pairing at only one value of momentum $\mathbf{q}_0$ is stable only
in a very limited phase space \cite{LOFF_Review}. Multiple plane wave
FFLO states are much more complicated and are currently under
investigation.

\begin{figure}
\includegraphics[width=3.4in,clip]{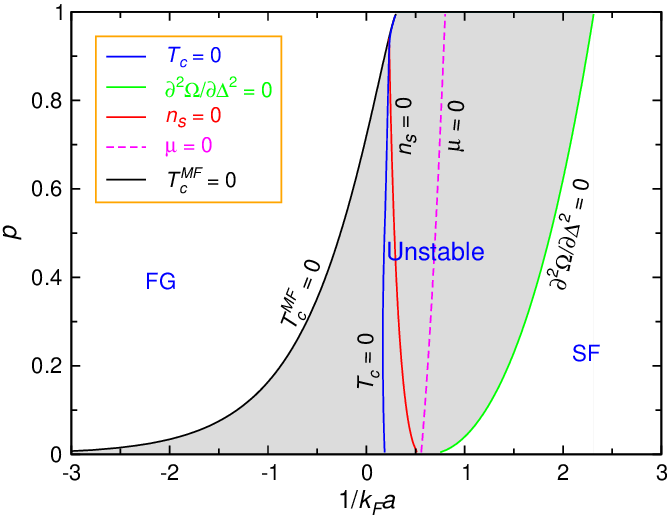}
\caption{Zero temperature phase diagram of a homogeneous Fermi gas as a
  function of pairing interaction characterized by $1/k_Fa$ and the
  polarization $p$.  On the left side of the (black) $T_c^{MF}=0$ curve,
  the system is a stable polarized Fermi gas (labeled by ``FG''). On
  the right side of the (green) $\p^2 \Omega/\p\Delta^2=0$ curve, the
  system is a stable polarized superfluid at low $T$, labeled by
  ``SF''. At $T=0$, the entire shaded region between these two curves
  is unstable against phase separation, and a stable polarized
  superfluid can exist only in the BEC regime.  The (red) $n_s=0$ curve
  and the (blue) $T_c=0 = 1/M^*$ curves are completely within the
  unstable regime. Here $k_F$ is the Fermi wavevector of a
  noninteracting two component Fermi gas of the same density $n$ without
  a population imbalance. All energies are measured in units of
  $E_F\equiv k_F^2/2m$. Throughout this paper, we use $k_0 = 80k_F$.}
\label{fig:PhaseT0}
\end{figure}

We set up the central issues of this paper by summarizing our previously
obtained \cite{Chien06} results on the zero temperature phase diagram in
the $p$--$1/k_Fa$ plane plotted in Fig.~\ref{fig:PhaseT0}. Here $p\equiv
\delta N/N$ is the polarization, $k_F$ is the Fermi wavevector of a
noninteracting Fermi gas of the same number density without a population
imbalance, $a$ is the two-body $s$-wave scattering length,
$N=N_\uparrow+N_\downarrow$ and $\delta N = N_\uparrow-N_\downarrow > 0$
are the total number and number difference, respectively. In the BCS
limit, our result for the boundary separating the stable polarized
normal Fermi gas and the unstable Sarma phase agrees with that in
Ref.~\cite{Pao}, but differs quantitatively from that in
Ref.~\cite{Radzihovsky}. On the BEC side, our result agrees with that of
Ref.~\cite{Radzihovsky} but not that of Ref.~\cite{Pao}.  Since there is
so much controversy in the literature, this paper, then, addresses a
particularly timely issue.

The differences between the various theoretical proposals for the phase
diagram stem from different conclusions concerning the stability of the
various phases.  There seems to be general agreement
\cite{Radzihovsky,Chien06,Tsinghua} about the two generic stability
requirements which were first articulated in Ref.~\cite{Pao} where it
was argued that the number susceptibility matrix must have positive
eigenvalues, and that the superfluid density must also be positive.  We
will show in detail here that in contrast to Ref.~\cite{Radzihovsky},
that the positivity of eigenvalues of the number susceptibility matrix
is equivalent to the positivity of the 2nd order partial derivative
$\frac{\p^2 \Omega}{\p\Delta^2}$, where $\Omega$ is the thermodynamic
potential, and $\Delta$ the fermionic excitation gap. The same
observation was very recently made in Ref.~\cite{Tsinghua}. In
comparison to these two equivalent conditions, we find that the
positivity requirement on the superfluid density is much less stringent.
This differs from Ref.~\cite{Tsinghua} but agrees with
Refs.~\cite{Pao,Radzihovsky}. In addition, we find that the requirement
that the pair mass be positive constitutes another nontrivial stability
condition, and may be more stringent than the positivity of $\frac{\p^2
\Omega}{\p\Delta^2}$ at low temperature and low imbalance in the BCS
regime.

Our theoretical formalism is described in Ref.~\cite{Chien06}, so we
will not repeat the details here. The system is composed of two spin
components of number $N_\uparrow$ and $N_\downarrow$ and of chemical
potential $\mu_\uparrow$ and $\mu_\downarrow$, respectively. They
interact via the attractive interaction $U_\mathbf{k,k'}= U \phik\phik'$
with $U<0$. Here we use a Gaussian cutoff $\phik = e^{-k^2/2k_0^2}$ with
$k_0$ sufficiently large (as appropriate for a short range
potential). The cutoff $\phik$ can be used to approximate a potential,
which includes but also generalizes the well studied contact
potential. Here $k_0$ is given by the inverse range of interaction, and
we take $k_0 = 80k_F$ in our numeric calculations throughout this
paper. This interaction is related to $1/k_Fa$ via $m/4\pi a = 1/U +
\sumk \phik^2/2\ek$, where $\ek = k^2/2m$.
The full fermion Green's function is given by
$G_{\uparrow,\downarrow}(K) = u_\mathbf{k}^2/(i\omega_n \pm h -\Ek) +
v_\mathbf{k}^2/(i\omega_n \pm h +\Ek) $, where $\Ek =
\sqrt{\xi_\mathbf{k}^2 +\Delta^2}$, $\xi_\mathbf{k} = \ek - \mu$,
$\mu=(\mu_\uparrow + \mu_\downarrow)/2$, $h=(\mu_\uparrow -
\mu_\downarrow)/2$ and $u_\mathbf{k}^2, v_\mathbf{k}^2 = (1\pm
\xi_\mathbf{k}/\Ek)/2$. As in Ref.~\cite{Chien06}, we set the volume
$V=1$, $\hbar=k_B = 1$, and $K\equiv (i\omega_n, \mathbf{k})$, $Q\equiv
(i\Omega_m, \mathbf{q})$, $\sum_K \equiv T\sum_n\sumk$, etc, where
$\omega_n (\Omega_m)$ is odd (even) Matsubara frequencies \cite{Fetter}.
Our Green's function is equivalent to presuming the BCS form for the
self-energy, $\Sigma_\sigma(K) = - \Delta^2 G_{\bar{\sigma}}^0
(-K)\phik^2$, where $\sigma=\uparrow,\downarrow$ and $\bar{\sigma} =
-\sigma$.
Here the bare Green's function
$G_{0,\sigma}^{-1} (K) = i \omega_{n} - \xi_{\mathbf{k},\sigma}$.
With these definitions we introduce the important pair susceptibility
\begin{equation}
\chi(Q) =  \mathop{\sum_K} G(K)G_0(Q-K)\varphi _{{\bf k}-{\bf q}/2} ^2,
\label{chi}
\end{equation}
which enters throughout the present paper.  We thus have $\sum_K
\Sigma_\sigma(K) G_\sigma(K) = -\Delta^2 \chi(0)$,

At finite $T$, we have four equations in the presence of population
imbalance. They are the gap ($\Delta$) equation, two number equations,
and the pseudogap ($\Delta_{pg}$) equation. At zero $T$ (where
$\Delta_{pg} =0$), our theory yields three equations and reduces to the
standard one-channel results in the literature.  More generally
$\Delta^2(T) = \Delta_{sc}^2 (T) + \Delta_{pg}^2 (T) $, where
$\Delta_{sc}$ is the superfluid order parameter.  Except for the number
difference equation, all equations can be written in the same form as in
the case with equal spin population \cite{ourreview}, provided one
replaces the Fermi distribution function $f(x)$ and its derivative
$f'(x)$ with the averages $\bar{f}(x)\equiv [f(x+h)+f(x-h)]/2$ and
$\bar{f}'(x)$. 
%
At $T \neq 0$, we must include the effects of noncondensed pairs via the
pseudogap term $\Delta_{pg}$.  These noncondensed pairs have been
ignored in previous work \cite{Machida,Tsinghua,latestStoof}, and are
found to be extremely important here.  They have an effective pair mass
$M^*$ which must necessarily be positive, thereby, adding another
condition to the requirements for phase stability.

At general $T$, our self consistent equations are
\bea 0&=&1+U\chi(0) = 1+ U\sumk \frac{1-
  2\bar{f}(E_{\mathbf{k}})}{2E_{\mathbf{k}}}\phik^2\,,
\label{eq:gap}
\eea
\begin{subequations}
\label{eq:neq}
\begin{eqnarray}
\label{eq:neqa}
n &=& 2\sum_\mathbf{k} \left[\vk^2 + \frac{\xi_\mathbf{k}}{\Ek}
  \bar{f}(\Ek)\right],
\\
pn &=& \sum_\mathbf{k} [f(\Ek-h)-f(\Ek+h)]
\label{eq:neqb}
\end{eqnarray}
\end{subequations}
\begin{equation}
  \Delta_{pg}^{2}\equiv - \sum_{Q\neq 0}t(Q) = \frac{1}{Z} \sumq b(\Omegaq)\,,
\label{eq:pg}
\end{equation}
Here $\chi(0)$ is the pair susceptibility at $Q=0$, and $\Omegaq =
q^2/2M^*$ is the pair dispersion. From Eq.~(\ref{eq:neqb}), we have
$h>\Delta$ at $T=0$ for any $p>0$. Solving
Eqs.~(\ref{eq:gap})-(\ref{eq:neq}) with $\Delta=0$ at $T=0$, we can
obtain the boundary between the polarized normal Fermi gas phase and the
unstable phase in Fig.~\ref{fig:PhaseT0}.

\section{Thermodynamic potential $\Omega$}

We first discuss the gap [Eq.~(\ref{eq:gap})] and number equations
[Eqs.~(\ref{eq:neq})], which govern the fermionic degrees of freedom,
and only later address pseudogap effects which appear at finite $T$
through Eq.~(\ref{eq:pg}). The gap and number equations can be obtained
from the fermionic part of the thermodynamic potential
$\Omega(\Delta,\mu,h)$ via partial derivatives with respect to $\Delta$,
$\mu$, and $h$, respectively. It should be noted that only the gap
equation corresponds to a vanishing first order derivative.

Instead of writing down the thermodynamic potential from the known
quasiparticle energy spectra \cite{Duan}, we calculate it via the energy
and entropy. We have 
\begin{subequations}
\bea
 K &\equiv& E-\mu_\uparrow N_\uparrow-\mu_\downarrow N_\downarrow =
E - \mu N - h \delta N, \\ 
\Omega &\equiv& E-TS-\mu N - h \delta N = K - TS \,,\\
F&\equiv&E-TS = \Omega + \mu N + h \delta N\,,
\label{eq:Omega}
\eea
\end{subequations}
and  
\be
N = -\left(\frac{\partial \Omega}{\partial \mu}\right)_{\Delta,h}, 
\qquad  \delta N =
-\left(\frac{\partial \Omega}{\partial h}\right)_{\Delta,\mu}\,.
\label{eq:Num}
\ee

The energy for each spin is given by
\bea
E_\sigma &=& \sum_K \left[ {\ek} + \frac{1}{2}
  \Sigma_\sigma(K)\right] G_\sigma(K) 
\,.
\eea 
%
Using the gap equation (\ref{eq:gap}), one can show that
\bea
E &=& \sum_\sigma E_\sigma = \sum_K \ek [G_\uparrow(K) +
  G_\downarrow(K)] -\Delta^2 \chi(0)\nonumber\\
&=& \sumk \left[ \xik -\Ek + 2\Ek \bar{f}(\Ek) \right] +  \mu N -
\frac{\Delta^2}{U} \,,\\
%
%
K &=& \sumk \left[ \xik -\Ek + 2\Ek \bar{f}(\Ek) \right] -h \delta N  -
\frac{\Delta^2}{U} \,.
\eea
%

The quasiparticle excitation energies are now given by $\pm \Ek + \sigma
h $, where $\sigma = \pm 1$ for spin down and up, respectively. The
entropy contains contributions from both spin species.
\bea 
S &=& -\sum_{\sigma=\pm1} \sumk \left[ f(-\Ek+\sigma h) \ln
  f(-\Ek+\sigma h) \right.   \nonumber\\ 
&&{} +\left. f(\Ek+\sigma h) \ln f(\Ek+\sigma h)\right]\,.
\eea

\be
\frac{\partial K}{\partial \Delta} = \sumk [2\Ek
  \bar{f}'(\Ek) + 2\bar{f}(\Ek) -1] \frac{\partial \Ek}{\partial
  \Delta} 
- h \frac{\partial \delta N}{\partial \Delta}
 - 2\frac{\Delta}{U}\nonumber
\ee
\be
\frac{\partial S} {\partial \Delta} = \frac{2}{T} \sumk  \bar{f}'(\Ek)
\Ek \frac{\partial \Ek}       {\partial \Delta}  - \frac{h}{T}
      \frac{\partial \delta N}{\partial \Delta}\nonumber
\ee

Finally, we obtain
\bea
\frac{\partial \Omega} {\partial \Delta} &= & 
\frac{\partial K} {\partial \Delta}  - T
\frac{\partial S} {\partial \Delta}   = \sumk
\left[ 2\bar{f}(\Ek) -1\right] \frac{\partial \delta N}{\partial
  \Delta} - 2\frac{\Delta}{U} \nonumber\\
&=&  - \frac{2\Delta}{U} \left[ 1 + U\chi(0)\right]\,,
\eea
where we have used the relationship $\partial \Ek/\partial \Delta =
  \Delta \phik^2/\Ek$. It is evident that when the gap equation is
  satisfied, $\partial \Omega/ \partial \Delta = 0$. Alternatively, one
  easily finds $\d F/\d\Delta =0$ at the same time. It is interesting to
  note that in Eq.~(\ref{eq:Omega}) the term $\mu N$ cancels that from
  $E$, whereas $h \delta N$ cancels that in $TS$. It is straightforward
  to verify Eqs.~(\ref{eq:Num}) are equivalent to the usual form of the
  number and number difference equations (\ref{eq:neqa}) and
  (\ref{eq:neqb}). Using Eqs.~(\ref{eq:Num}), 

When the gap equation is satisfied, we readily obtain
\begin{subequations}
\bea
\frac{\partial^2 \Omega} {\partial \Delta^2}  &=& -2\left[\frac{1}{U} +
  \chi(0)\right]
-2\Delta \frac{\partial \chi(0)} {\partial \Delta}\nonumber\\
&=& 2 \sumk \frac{\Delta^2 \phik^4}{\Ek^2}
\left[\frac{1-2\bar{f}(\Ek)}{2\Ek} + \bar{f}'(\Ek) \right] .
\label{eq:d2Delta}
\eea

Additional second order partial derivatives of interest are 
\bea
\frac{\partial^2 \Omega} {\partial \mu \partial \Delta}
&=& - 2 \sumk
\frac{\Delta \ek \phik^2 }{\Ek^2} \left[\frac{1-2\bar{f}(\Ek)}{2\Ek} +
  \bar{f}'(\Ek) \right] \nonumber\\
 &=& - \frac{\partial N} {\partial \Delta}\,,\\  
\frac{\partial^2 \Omega} {\partial \mu \partial h} 
&=&   \sumk  \frac{\ek}{\Ek}   \left[ f'(\Ek-h) - f'(\Ek+h) \right]
\nonumber \\  
&=& - \frac{\partial  N} {\partial h} = -
 \frac{\partial \delta N} {\partial \mu} \,,\\  
\frac{\partial^2 \Omega} {\partial  \Delta\partial h}
&=&   -\sumk 
\frac{\Delta  \phik^2 }{\Ek} \left[ f'(\Ek-h) - f'(\Ek+h)\right] \nonumber\\
& =&
-\frac{\partial \delta N} {\partial \Delta}\,,\\  
%
%
\frac{\partial^2 \Omega} {\partial \mu^2} &=& -\frac{\partial  N}
     {\partial \mu}  =  2\sumk 
\bigg\{  \bar{f}'(\Ek)  \nonumber\\
&&{}  -
\frac{\Delta^2\phik^2}{\Ek^2} \left[ \frac{1-2\bar{f}(\Ek)}{2\Ek} +
  \bar{f}'(\Ek) \right]     \bigg\}\,,\\
\frac{\partial^2 \Omega} {\partial h^2} &=& - \frac{\partial \delta N}
     {\partial h}  = 2\sumk \bar{f}'(\Ek) \,.
\eea
\end{subequations}

\section{Stability conditions}
\label{sec:Stability}

\subsection{Stability of the polarized normal phase on the BCS side}
\label{subsec:FG}

The condition $\p\Omega/\p\Delta=0$ admits a trivial solution
$\Delta=0$.  Referring now to Fig.~\ref{fig:PhaseT0}, we see that on the
\emph{left side} of the boundary $T_c^{MF}=0$ between the $\Delta=0$ and
the paired phase, the quantity $\chi(0)$ is a function of $\mu$ and $h$
only. Here we have $U^{-1}+\chi(0) < 0$, independent of $1/k_Fa$.  It is
easy to show that the expression inside the square brackets of the
second line in Eq.~(\ref{eq:d2Delta}) vanishes when $\Ek <
h$. Therefore, when $\Delta=0$ but $1+U\chi(0)\neq 0$, we always have
\be
\frac{\partial^2 \Omega} {\partial \Delta^2}  = -2\left[\frac{1}{U} +
  \chi(0)\right]  > 0\,.
\label{eq:d2D_FG}
\ee
We will argue later that the positivity of $\frac{\partial^2 \Omega}
{\partial \Delta^2}$ is associated with stability against phase
separation.  From this and the above inequality, we can conclude that
the homogeneous polarized Fermi gas phase in Fig.~\ref{fig:PhaseT0} is
stable with respect to phase separation.

\subsection{Second order total derivative $\displaystyle \frac{\d^2F
  } {\d \Delta^2} > 0$ -- Stability of pairing versus normal Fermi  gas
  for fixed $N$ and $\delta N$}

Next we investigate the total derivative $\displaystyle \frac{\d^2 F }
  {\d \Delta^2} $, for a fixed particle number system. This is
  equivalent to confining the equations in the hyperplane defined by the
  two number equations in the parameter space spanned by $\Delta, \mu$
  and $h$. The constraints imposed by the number equations
  (\ref{eq:neq}) exclude the possibility of phase separation or any
  spatial variation of the order parameter.  We will show that $\d^2
  F/\d\Delta^2 >0 $ provides a stability condition for superfluidity
  \textit{vis a vis} a polarized normal Fermi gas phase. This is
  equivalent to the statement that the free energy reaches a minimum
  when a nontrivial solution ($\Delta\ne 0$) of our equation set is
  found.

From the number equations, we have 
\bea
0 &=& \frac{\d N}{\d\Delta} = \frac{\p N}{\p\Delta} + \frac{\p N}{\p \mu}
\frac{\d \mu}{\d \Delta} + \frac{\p N}{\p h}
\frac{\d h}{\d \Delta}\,, \\
0 &=& \frac{\d \delta N}{\d\Delta} = \frac{\p \delta  N}{\p\Delta} + \frac{\p
 \delta   N}{\p \mu} \frac{\d \mu}{\d \Delta} + \frac{\d \delta  N}{\d h}
\frac{\d h}{\d \Delta} \,,
\eea
so that
\be
\begin{pmatrix}\displaystyle \frac{\d \mu}{\d
    \Delta} \\ \\\displaystyle \frac{\d h}{\d \Delta}
\end{pmatrix} 
=\frac{-1}{\displaystyle \frac{\p (N,\delta N)}{\p (\mu,h)}}
    \begin{pmatrix}\displaystyle \frac{\p (N,\delta N)}{\p 
    (\Delta,h)} \\ \\\displaystyle \frac{\p (N,\delta N)}{\p (\mu,\Delta)}
\end{pmatrix} \,.
\ee
Using the chain rule, 
\be
\frac{\d \Omega}{\d\Delta} = \frac{\p \Omega}{\p\Delta} + \frac{\p
  \Omega}{\p\mu}\frac{\d\mu}{\d\Delta} + \frac{\p
  \Omega}{\p h}\frac{\d h}{\d\Delta} \,,
\label{eq:chain}
\ee
finally, we obtain
\bea
\frac{\d^2 F}{\d \Delta^2} 
&=& \frac{\d^2 \Omega}{\d \Delta^2} + N \frac{\d^2\mu}{\d\Delta^2} +
\delta N  \frac{\d^2 h}{\d\Delta^2} \nonumber\\
&=&  \frac{\p^2 \Omega}{\p \Delta^2}  +  \frac{\p^2 \Omega}{\p \Delta\p\mu} \frac{\d \mu}{\d \Delta} 
+  \frac{\p^2\Omega}{\p \Delta\p h} \frac{\d h}{\d \Delta} \,,
  \nonumber\\
\eea
where we have used Eq.~(\ref{eq:chain}) to simplify this expression.

\begin{figure}
\includegraphics[width=3.4in,clip]{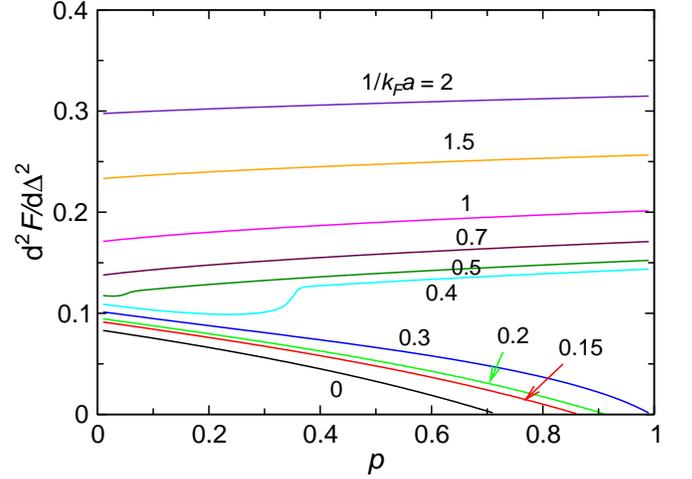}
\caption{Behavior of $\d^2 F/\d\Delta^2$ at $T=0$ as a function of
  polarization $p$ for various interaction strength $1/k_Fa$, from BCS
  to BEC, as labeled in the figure. From top to bottom, $1/k_Fa$
  decreases monotonically. $\d^2 F/\d\Delta^2$ remains positive on the
  right side of the $T_c^{MF}=0$ boundary in Fig.~\ref{fig:PhaseT0}, and
  vanishes on this boundary.}
\label{fig:d2WdD2}
\end{figure}

We plot the total derivative $\d^2 F/\d\Delta^2$ at $T=0$ as a function
of polarization in Fig.~\ref{fig:d2WdD2}. This figure demonstrates that
for all regions on the right side of the $T_c^{MF}=0$ line in
Fig.~\ref{fig:PhaseT0}, (where the solution of the gap equation with
fixed particle number exists), $\d^2 F/\d\Delta^2$ is always positive.
It vanishes precisely at this boundary, that is, on the line $T_c^{MF}
=0$, which separates the paired phase and the polarized normal Fermi gas
phase. For definiteness, at unitarity, $\d^2 F/\d\Delta^2$ vanishes at
$p=0.72$.  Above this polarization the system is stable as a normal,
unpaired Fermi gas. Extrapolation of $\d^2 F/\d\Delta^2$ to the normal
Fermi gas side of this line would lead to a negative value, indicating
that the paired state is not stable there.

In summary, when a solution of the self-consistent equations
(\ref{eq:gap})-(\ref{eq:pg}) can be found, a paired phase is expected to
be more stable than the unpaired Fermi gas phase, both at zero and
finite $T$. This result agrees with what we obtained in the previous
subsection \ref{subsec:FG}, but addresses this issue from a different
perspective. Here we consider the stability of the paired phase whereas
in Section \ref{subsec:FG} we addressed that of the unpaired Fermi gas.
We can turn this around to conclude that \emph{the positivity of $\d^2
F/\d\Delta^2$ does not provide an extra constraint on the stability of
the polarized superfluid phase.}

\subsection{Number susceptibility and stability of homogeneous polarized
  superfluid against phase separation}

In this section, we consider the number susceptibility with respect to
variation of the chemical potentials in equilibrium. It is expected on
physical grounds \cite{Pao} that both eigenvalues of the matrix $ \p
N_\sigma/\p \mu_{\sigma'} $ must be positive in order for the system to
be stable, against phase separation. It should be obvious that when
$\Delta$ is held fixed, the eigenvalues of the matrix $ \p N_\sigma/\p
\mu_{\sigma'} $ are always positive. However, when $\Delta$ is treated
as an implicit function of $\mu$ and $h$ as defined by the gap equation
(\ref{eq:gap}), the eigenvalues may change sign.

For simplicity in notation, we use partial derivative $\p/\p x$ when
$\Delta$, $\mu$ and $h$ are all treated as independent variables, as in
the previous section. When $\Delta$ is regarded as a function of $\mu$
and $h$, we define
\be
\frac{\D}{\D x} \equiv \frac{\p }{\p x} + \frac{\p \Delta}{\p x}
\frac{\p}{\p \Delta}, \qquad (x = \mu, h) \,.
\ee

With the relationships 
$N = N_\uparrow + N_\downarrow$, $\delta N = N_\uparrow - N_\downarrow$,
and $\frac{\D} {\D \mu_\uparrow} = \frac{1}{2} \left( \frac{\D} {\D
  \mu} +  \frac{\D} {\D h} \right)$,  $\frac{\D} {\D \mu_\downarrow} =
\frac{1}{2} \left( \frac{\D} {\D  \mu} -  \frac{\D} {\D h} \right)$ , we
can easily find 
\bea
\begin{pmatrix}\displaystyle
\frac{\D N_\uparrow}{\D \mu_\uparrow} &\displaystyle \frac{\D
  N_\uparrow}{\D \mu_\downarrow} \\ \\\displaystyle 
\frac{\D  N_\downarrow}{\D \mu_\uparrow} &\displaystyle \frac{\D
  N_\downarrow}{\D\mu_\downarrow }\end{pmatrix} 
&=& \frac{1}{2} A
  \begin{pmatrix}\displaystyle 
\frac{\D N}{\D \mu} &\displaystyle \frac{\D N}{\D h} \\ \\\displaystyle
\frac{\D \delta N}{\D \mu} &\displaystyle \frac{\D \delta N}{\D h} 
\end{pmatrix}  A \,.
\eea
Here $A=A^{-1}=\frac{1}{\sqrt{2}} \begin{pmatrix}1 & 1  \\ 1 &
-1\end{pmatrix}$ is a unitary transformation matrix. Therefore, the
eigenvalues of the matrix 
\be
M_{2\times2}=\begin{pmatrix}\displaystyle \frac{\D N}{\D
\mu} &\displaystyle \frac{\D N}{\D h} \\ \\\displaystyle \frac{\D \delta
N}{\D \mu} &\displaystyle \frac{\D \delta N}{\D h}
\end{pmatrix} 
\ee
are also required to be positive.

From the gap equation, we readily obtain
\bea
\frac{\p\Delta}{\p \mu} &=& -\frac{\p \chi/\p \mu} {\p\chi/\p\Delta} = -
\frac{\p^2 \Omega}{\p\Delta\p\mu}/\frac{\p^2 \Omega}{\p\Delta^2} ,\\
\frac{\p\Delta}{\p h} &=& -\frac{\p \chi/\p h} {\p\chi/\p\Delta} = -
\frac{\p^2 \Omega}{\p\Delta\p h}/\frac{\p^2 \Omega}{\p\Delta^2} \,.
\eea
It should be noted that this expression is valid only when the gap
equation (\ref{eq:gap}) is satisfied, i.e., 
on the right side of the $T_c^{MF}=0$ boundary in Fig.~\ref{fig:PhaseT0}. On
the left side, we have the inequality of Eq.~(\ref{eq:d2D_FG}).

Finally, we have
\begin{subequations}
\label{eq:2x2_elements}
\bea
\frac{\D N}{\D\mu} & =& 
- \frac{\p^2\Omega}{\p\mu^2} +
\left(\frac{\p^2\Omega}{\p\mu\p\Delta}\right)^2 /\frac{\p^2
  \Omega}{\p\Delta^2} \,,\\
\frac{\D N}{\D h} & =& 
- \frac{\p^2\Omega}{\p\mu\p h} +
\frac{\p^2\Omega}{\p\mu\p\Delta} \frac{\p^2\Omega}{\p\Delta\p h } /\frac{\p^2
  \Omega}{\p\Delta^2} = \frac{\D\delta N}{\D \mu}\,,\\
\frac{\D\delta N}{\D h} & =& 
- \frac{\p^2\Omega}{\p h^2} +
\left( \frac{\p^2\Omega}{\p\Delta\p h }\right)^2 /\frac{\p^2
  \Omega}{\p\Delta^2} \,.
\eea
\end{subequations}

Therefore, the eigenvalues are given by
\bea
\label{eq:M2x2_EV}
\lambda_\pm &=& \frac{\Tr(M_{2\times2})\pm \sqrt{\Tr(M_{2\times2})^2 - 4
    \det(M_{2\times2})}}{2} \,,
\eea
where
$\Tr(M_{2\times2})=\frac{\D N}{\D\mu} + \frac{\D \delta N}{\D
  h}$ and $\det(M_{2\times2}) = \frac{\D N}{\D\mu} \frac{\D
  \delta N}{\D  h} - \left(\frac{\D N}{\D h}\right)^2 $ are the trace
and determinant, respectively.

Since $\frac{\p^2\Omega}{\p \Delta^2}$ appears in the denominator of the
expressions in Eqs.~(\ref{eq:2x2_elements}), each element of $M_{2\times2}$
will change sign when $\frac{\p^2\Omega}{\p \Delta^2}$ approaches zero
and changes sign. Under this circumstance,
$\left[\det(M_{2\times2})\right]^{-1}$ approaches zero and changes
sign. Our numerics shows that only one of the two eigenvalues in
Eq.~(\ref{eq:M2x2_EV}) changes sign. This roughly corresponds to
$\frac{\D \delta N}{\D h}$. Therefore, \emph{the stability condition
that the eigenvalues of the number susceptibility matrix be positive is
equivalent to}
\be 
\label{eq:d2D}
\frac{\p^2\Omega}{\p \Delta^2}>0 \,, \qquad \mbox{when $\quad 1+U\chi(0)=0$}\,.
\ee

\begin{figure*}
\includegraphics[width=6.in,clip]{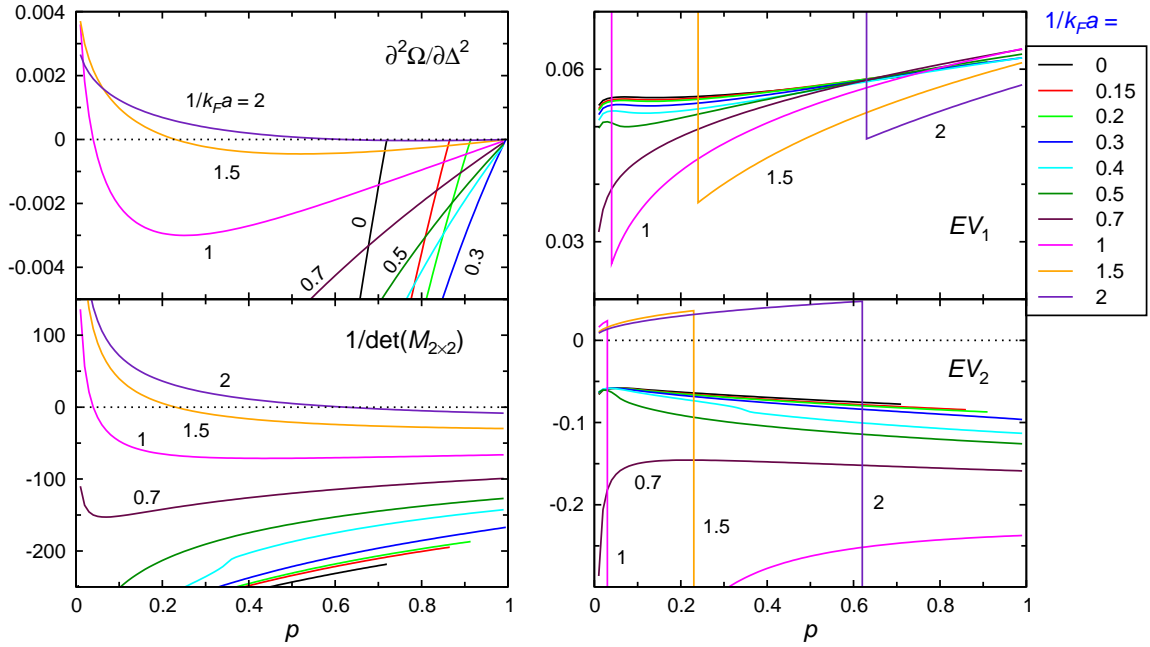}
\caption{Behavior of $\p^2 \Omega/\p\Delta^2$, the inverse determinant
  $1/\det(M_{2\times2})$ and the eigenvalues of the number
  susceptibility matrix $M_{2\times2}$ at $T=0$ as a function of
  polarization $p$ for various interaction strength $1/k_Fa$, from
  unitarity to BEC, as labeled in the figure. The first eigenvalue,
  $EV_1\equiv \lambda_+$, remains positive, roughly corresponding to $\D
  N/\D\mu$. The sign change of $\p^2 \Omega/\p\Delta^2$ and the second
  eigenvalue, $EV_2\equiv \lambda_-$, and thus $1/\det(M_{2\times2})$
  occurs simultaneously, for $1/k_Fa=1, 1.5$ and 2 on the BEC side of the
  Feshbach resonance, along the $\p^2 \Omega/\p\Delta^2=0$ phase
  boundary in Fig.~\ref{fig:PhaseT0}. }
\label{fig:EV}
\end{figure*}

This condition has been argued in the literature to correspond to the
condition for phase separation. Indeed, this condition says when $\mu$
and $h$ are held fixed, a stable homogeneous superfluid solution should
always minimize the thermodynamic potential. Otherwise, the system tends
to phase separate into a region with smaller $\Delta$ and another region
with larger $\Delta$, both of which give lower $\Omega$. In such a case,
the number density in each region is \emph{not} fixed. Such a phase
separation is generally believed to be realized by a one-component Fermi
gas physically adjacent to an unpolarized superfluid.  We must have
$\Delta > h$ in the unpolarized superfluid region. If we take into
account the possibility of condensation at $\mathbf{q}_0\neq 0$, this
form of microscopic phase separation may not be the only way to address
the instability. One may have a more homogeneous state as well such as
found in FFLO-like phases.

The behavior of $\p^2 \Omega/\p\Delta^2$, the inverse determinant
$1/\det(M_{2\times2}) = 1/(\lambda_+\lambda_-)$ and the eigenvalues of
the number susceptibility matrix $M_{2\times2}$ at $T=0$ are all plotted
in Fig.~\ref{fig:EV} as a function of polarization $p$ for various
interaction strengths $1/k_Fa$. From Eqs.~(\ref{eq:2x2_elements}), it is
evident that all elements of the matrix $M_{2\times2}$ change sign where
$\p^2 \Omega/\p\Delta^2 =0$. Nevertheless, we notice that one of the two
eigenvalues, $EV_1\equiv \lambda_+$, is always positive. This roughly
corresponds to $\D N/\D \mu $. However, the second eigenvalue,
$EV_2\equiv \lambda_-$, does change sign for $1/k_Fa = 1, 1.5,$ and 2,
exactly where $\p^2 \Omega/\p\Delta^2 = 0$.  Moreover, the fact that
$\p^2\Omega/\p\Delta^2$ appears in the denominator in
Eqs.~(\ref{eq:2x2_elements}) is manifested by the jump in $\lambda_\pm$
upon the sign change.

If one extrapolates $\p^2 \Omega/\p\Delta^2$ in the upper left panel,
for $1/k_Fa \lesssim 0.3$, one would see that this partial derivative
becomes positive above the $T_c^{MF}=0$ line in
Fig.~\ref{fig:PhaseT0}. This reconfirms that the polarized normal Fermi
gas phase in Fig.~\ref{fig:PhaseT0} is stable. One can see thus that the
$T_c^{MF}=0$ provides another $\p^2 \Omega/\p\Delta^2=0$ (at
$\Delta\rightarrow 0$) line, but this time on the BCS side of the
resonance. Note the contrasting behavior of $\p^2 \Omega/\p\Delta^2$ and
$\d^2 F/\d\Delta^2$ across the $T_c^{MF}=0$ line.

At this stage it is useful to compare with the literature.  It is argued
in Ref.~\cite{Radzihovsky} that the positivity of eigenvalues of the
matrix $M_{2\times2}$ is a necessary but not sufficient condition for
stability. Moreover, in this previous work, the positivity of
eigenvalues of the number susceptibility matrix was found to be
different from the condition contained in Eq.~(\ref{eq:d2D}).  By
contrast, here we find that these two conditions are indeed equivalent.
The same observation was made in Ref.~\cite{Tsinghua}, although only the
matrix element $\D\delta N/\D h$ was addressed.

In other recent work \cite{Duan,Mueller}, a numerical optimization
procedure was invoked in which the entire landscape of $\Omega$ as a
function of $\Delta$ for fixed $\mu$ and $h$ was used to find the most
stable phase. It was argued that $\p^2 \Omega/\p\Delta^2 = 0$ often gave
multiple solutions, sometimes corresponding to a local maximum. In such
cases, the $\Delta=0$ solution was sometimes claimed to be more stable,
since it minimizes $\Omega$ globally. 

These results should be contrasted with the present calculations.
Except for the trivial $\Delta=0$ solution, we find there is at most one
solution to our set of equations at either zero or finite $T$.  We argue
that the optimization method is in principle valid, but it requires that
the system be in chemical equilibrium with an infinitely large particle
reservoir so that $\mu$ and $h$ are unchanged before and after a
possible phase separation. Because the atomic Fermi gases constitute a
finite system, $\mu$ and $h$ will be different before and after phase
separation; they, thus, no longer satisfy the same particle number
constraint. In this way we conclude that one cannot use a single
$\Omega(\Delta)$ curve to find the solution for the stable phase for a
given polarization at fixed particle number.

\subsection{Superfluid density}

For a superfluid state to be stable, another obvious condition
\cite{Pao} is that the superfluid density must be positive. Using linear
response theory, one can calculate the superfluid density via the
response of the system to an external (fictitious) vector potential, as
if it were a charged superconductor. In this way one computes the the
current-current correlation functions compatible with our $T$-matrix
approximation \cite{ourreview}, as shown in Ref.~\cite{Kosztin2}. Here,
however, we need to keep track of the population imbalance. Without
showing the details, our superfluid density is given by
\bea
\label{eq:ns}
\frac{n_s}{m} &=& \frac{2}{3}\sumk \frac{\Delta_{sc}^2}{\Ek^2} \left[
  \frac{1-2\bar{f}(\Ek)}{2\Ek} +  \bar{f}'(\Ek) \right] \nonumber\\
&&\times {}
     \left[\phik^2(\nabla\xik)^2 - \frac{1}{4} (\nabla\xik^2)\cdot
       (\nabla\phik^2)\right] \nonumber\\
&=&  n_s^{MF} \left(\frac{\Delta_{sc}^2}{\Delta^2}\right)  > 0 \,,
\eea
where $n_s^{MF}$ is an artificial construct corresponding
to the superfluid density in a BCS-like strictly mean-field
 treatment.  The boundary $n_s=0$ at $T=0$ is given by the red line in
 Fig.~\ref{fig:PhaseT0}.  This line is completely in the fermionic
 regime ($\mu>0$) and within the phase boundary set by $T_c^{MF}=0$ and
 $\p^2 \Omega/\p\Delta^2 = 0$. The superfluid density is positive on the
 right side of this line, and negative otherwise. We may summarize. The
 positivity of $n_s$ is a much weaker constraint than that given by
 Eq.~(\ref{eq:d2D}), corresponding to the positive definiteness of the
 number susceptibility matrix. This is different from
 Ref.~\cite{Tsinghua}.

It should be noted that at finite $T>T_c$, $n_s$ vanishes
  identically. However.  $n_s^{MF}$ may remain finite as long as the
  mean-field equations (\ref{eq:gap}) and (\ref{eq:neq}) are
  satisfied. In this case, $n_s^{MF}$ may change sign.

Our numerical result shows that the $n_s=0$ line shifts to the upper
left in the $p-1/k_Fa$ phase diagram as $T$ increases from zero, and
disappears completely slightly below $T/T_F < 0.2$. Based on
Fig.~\ref{fig:PhaseT0}, we conclude that $n_s$ remains non-negative for
$1/k_Fa > 0.5$.

\begin{figure*}
\includegraphics[width=6.5in,clip]{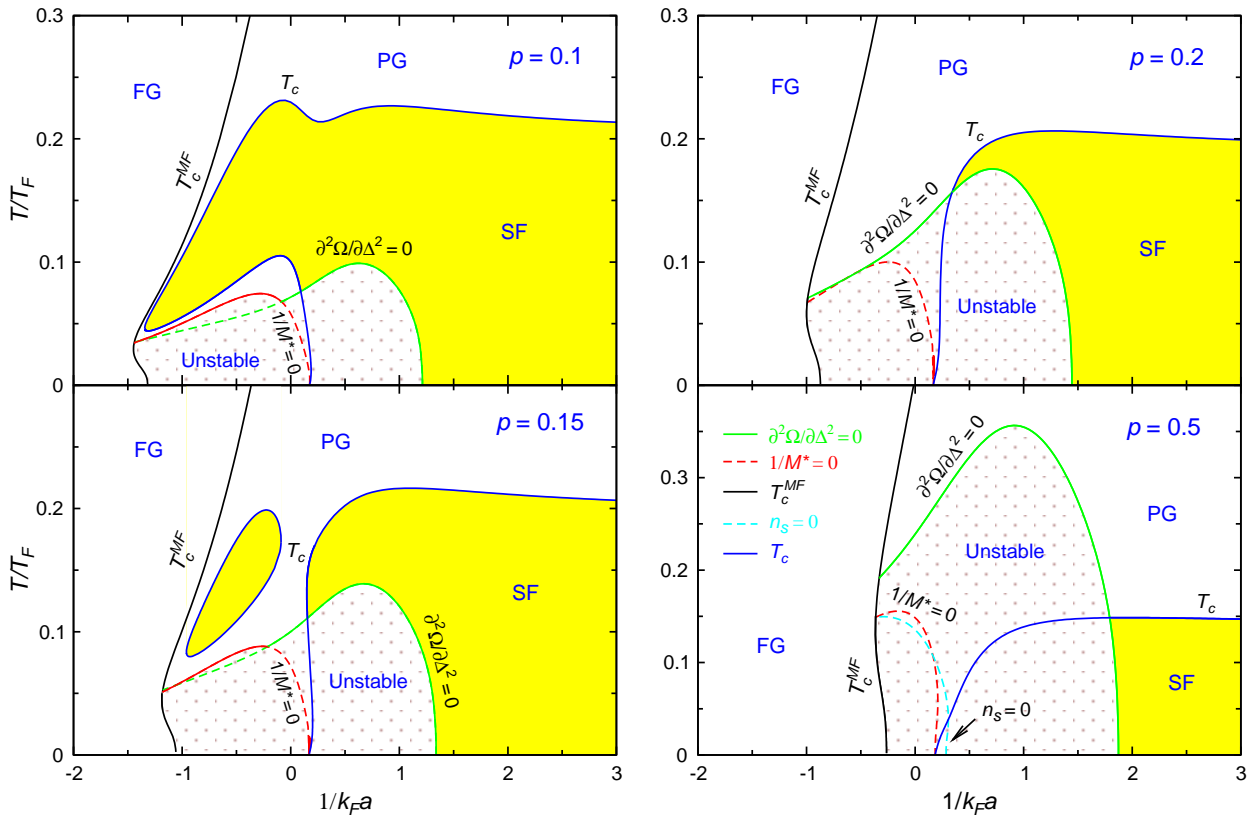}
\caption{Phase diagram in the $T-1/k_Fa$ plane for representative values
  of $p=0.1$, 0.15, 0.2, and 0.5 (as labeled). Shown are $T_c$ (blue
  line), $T_c^{MF}$ (black), and the instability boundaries defined by
  $\p^2\Omega/\p\Delta^2=0$ (green), and $1/M^*=0$ (red line),
  respectively. The yellow shaded area represents stable superfluid
  (labeled by ``SF''), the dotted region is unstable. The white open
  space on the right of the $T_c^{MF}$ line represents a stable
  pseudogap phase (``PG''), whereas on its left lies the stable unpaired
  Fermi gas state (``FG''). At $p\approx 0.14$, the superfluid region
  splits into two, and the small closed region shrinks with increasing
  $p$, and disappears at $p\approx 0.18$. For $p=0.1$ and 0.15, the two
  instability lines intersect with each other, whereas for $p=0.2$ and
  0.5, $1/M^*=0$ is completely inside the unstable region defined by
  $\p^2\Omega/\p\Delta^2 <0$. The (segment of each) stability line
  appear as a dashed line when it is inside an unstable region defined
  by the other stability condition. We also show the line on which $n_s=0$
  (cyan dashed curve) but for $p=0.5$ only since it is always inside the
  unstable region.  }
\label{fig:Tc}
\end{figure*}

\subsection{Positivity of the effective pair mass}

The effective pair mass can be obtained by Taylor expanding the
important pair susceptibility $\chi(Q)$ in the form
\be
\chi(Q)-\chi(0) = Z \left(i\Omega_m - \frac{q^2}{2M^*}\right) + \ldots . 
\ee

Most theoretical work in the literature on polarized Fermi gases is
based on the ground state \cite{Pao}, or at most on a BCS-like
mean-field treatment at finite $T$ \cite{Tsinghua,latestStoof}, which does not
include noncondensed pairs.  Here, by contrast, we include these
noncondensed pairs ($\Delta_{pg}^2 \neq 0$) or pseudogap effects
\cite{ChenPRL98} which are reflected in the pair propagator (or
$T$-matrix) and its associated dispersion.  As a result, we require that
the pair mass be positive.  This condition appears as another line in
the phase diagram of Fig.~\ref{fig:PhaseT0}.  At $T=0$, this would
correspond to the blue $T_c=0$ line, which, at this temperature, is
completely inside the unstable regime and does not provide an extra
constraint. However, at finite $T$, this line moves considerably to the
BCS side, especially at low $p$. As will be shown below, at low $T$ and
low $p$, this imposes a more stringent constraint than
Eq.~(\ref{eq:d2D}) in the BCS regime.

\section{Temperature dependence of the phase diagram}

\begin{figure}
\includegraphics[width=3.4in,clip]{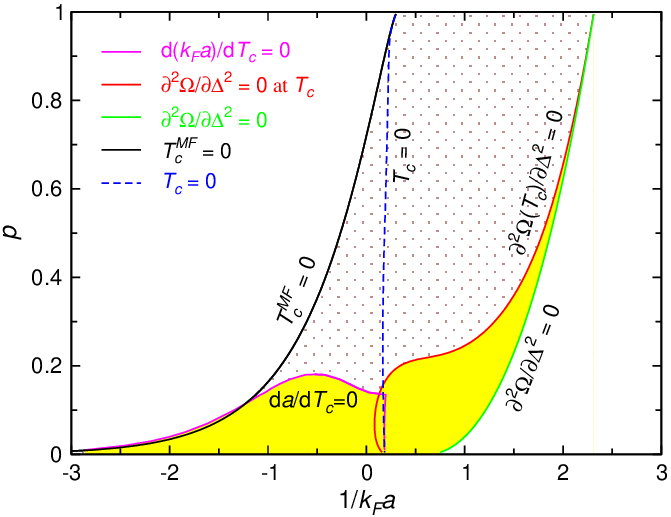}
\caption{Phase diagram in the $p-1/k_Fa$ plane, showing where
intermediate temperature superfluidity (shaded region) exists. The
$T_c^{MF}=0$, $T_c=0$, and zero temperature $\p^2 \Omega/\p\Delta^2=0$
lines are the same as in Fig.~\ref{fig:PhaseT0}.  The red curves
represent where $\p^2 \Omega/\p\Delta^2=0$ at corresponding $T_c$. The
line defined by $\d (1/k_Fa)/\d T_c =0$ is given by the turning points
$(p,1/k_Fa)$ where $1/k_Fa$ reaches a local extremum as a function of
$T_c$.  At $T=0$, the entire region between the $T_c^{MF}=0$ and the
$\p^2 \Omega/\p\Delta^2=0$ lines is unstable against phase
separation. However, a stable polarized, intermediate temperature
superfluid phase exists for the yellow shaded region enclosed by the
(orange) $\d (k_Fa)/\d T_c=0$ curve and the $p=0$ axis, or by the (red)
$\p^2\Omega(T=T_c)/\p\Delta^2=0$ and (green) $\p^2 \Omega/\p\Delta^2=0$
curves. Superfluidity in the dotted region is unstable at any
temperature.}
\label{fig:PhaseT1}
\end{figure}

Since superfluidity is essentially a finite temperature phenomenon, it
is important to assess the stability of the various superfluid phases at
$ T \neq 0$.  This is where experiments are performed.  Our theory
allows us to calculate the superfluid transition temperature $T_c$ in
the presence of the pseudogap effects associated with incoherent finite
center-of-mass momentum pair excitations.  We will see that quite
systematically, superfluid phases which are unstable at zero
temperature, may become stable at finite $T$.  As discussed in
Ref.~\cite{Chien06} these intermediate temperature superfluids will
behave somewhat differently depending on whether they occur to the right
or the left of the $T_c =0$ quantum critical point line in
Fig.~\ref{fig:PhaseT0}.

\begin{figure}
\includegraphics[width=3.3in,clip]{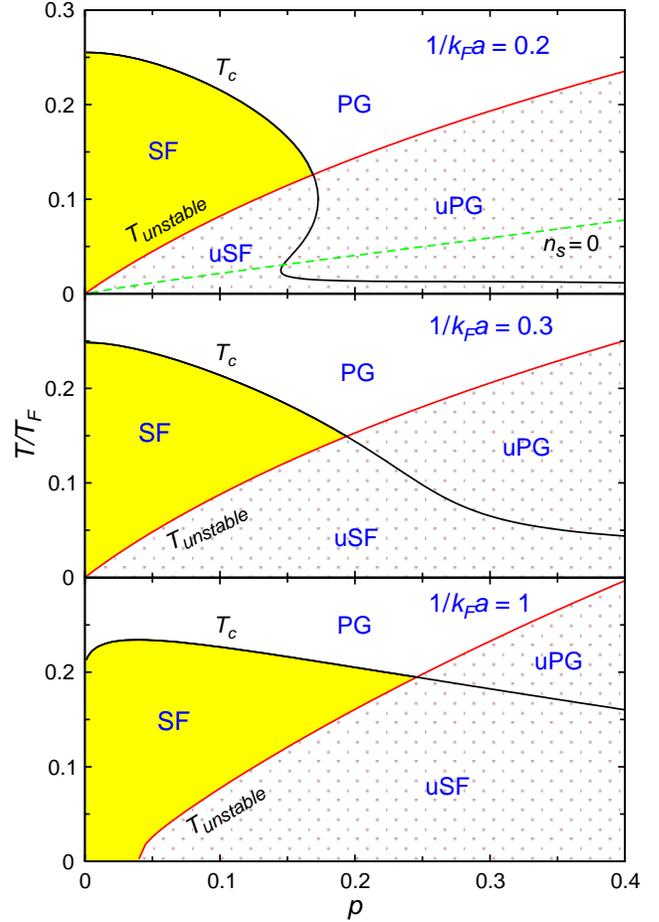}
\caption{Behavior of $T_c$ and the instability phase boundary on the BEC
  side. $T_{unstable}$ is defined by $\p^2 \Omega/\p\Delta^2=0$ in the
  $T$-$p$ plane, for $1/k_Fa = 0.2$, 0.3, and 1, from top to bottom. For
  each value of $1/k_Fa$, the phase diagram is composed of four
  different phases, separated by the solid lines: stable superfluid
  (SF), unstable superfluid (uSF), stable pseudogap (PG), and unstable
  pseudogap (uPG) phases, as labeled in the figure.  The yellow shaded
  region, on the left of the (black) $T_c$ and (red) $T_{unstable}$
  curves, e.g., the shaded area for $1/k_Fa = 0.2$, is a stable
  polarized superfluid. The superfluid solution below these two curves
  is unstable. Above these two curves, there exists a stable pseudogap
  phase. The unstable phases will disappear when $1/k_Fa > 2.3$. The
  pair mass is always positive for $1/k_Fa > 0.2$. We also show for
  $1/k_Fa = 0.2$ the $n_s=0$ line, which is completely within the
  unstable region defined by Eq.~(\ref{eq:d2D}). $n_s$ does not change
  sign at any $T$ and $p$ for $1/k_Fa >0.5$.}
\label{fig:T_unstable}
\end{figure}

In Fig.~\ref{fig:Tc} we present a slice of the finite temperature phase
diagram, plotted as characteristic temperature versus $1/k_Fa$, at four
representative polarizations, $p=0.1$, 0.15, 0.2 and 0.5.  The lines in
the figures correspond to $T_c$ and the phase boundaries defined by the
instability conditions, including where pairs are no longer found to be
stable (via $1/M^* <0$).  The shaded regions indicate where stable
superfluidity is found. The dotted regions indicate where either the
phase separation stability criterion (given by $\frac{\p^2
\Omega}{\p\Delta^2} >0$) is violated, or the effective pair mass becomes
negative.  It is important to note that on the BCS side of resonance
(depending on the polarization) there may be two values of $T_c$ for
each value of $1/k_Fa$.  The meaning of these two $T_c$'s was discussed
in earlier work \cite{Chien06}.  This case applies to the unitary
regime, as should be seen in the figure.  Below the lower $T_c$ we find
no solution to our set of four equations
(\ref{eq:gap})-(\ref{eq:pg}). In this way the system cannot support
superfluidity. More concretely, the effective pair chemical potential
become negative below the lower $T_c$. (At even lower $T$, the effective
pair mass becomes negative). Above the upper $T_c$, superfluidity is
destroyed in the usual way by finite temperature effects and the system
becomes a normal Fermi gas, which is far from ideal.  Here there are
still strong pairing correlations giving rise to a pseudogap in the
fermionic excitation spectrum.

By contrast as we move towards the BEC regime we find only one $T_c$,
but the stability of the superfluid at low $T$ is cut off because of the
negativity of $\frac{\p^2 \Omega}{\p\Delta^2}$.  Finally, sufficiently
deep into the BEC regime, a stable superfluid persists for all
temperatures (below $T_c$) including $T=0$.

These results are summarized in the form of a finite temperature phase
diagram in Fig.~\ref{fig:PhaseT1}.  The shaded region indicates where
intermediate temperature superfluidity occurs, while the dotted region
corresponds to an unstable superfluid, presumably a LOFF-like or
heterogeneous phase. Because there is a regime where two $T_c's$ are
evident, as seen in Fig.~\ref{fig:Tc}, we indicate the upper boundary of
the shaded region by a ``turning point condition": $\d (1/k_Fa) / \d T_c
=0$.

The condition given by Eq.~(\ref{eq:d2D}) at $T_c$ is represented by the
red curve in Fig.~\ref{fig:PhaseT1}. Note that $T_c$ self-consistently
along this line.

\begin{figure}
\includegraphics[width=3.3in,clip]{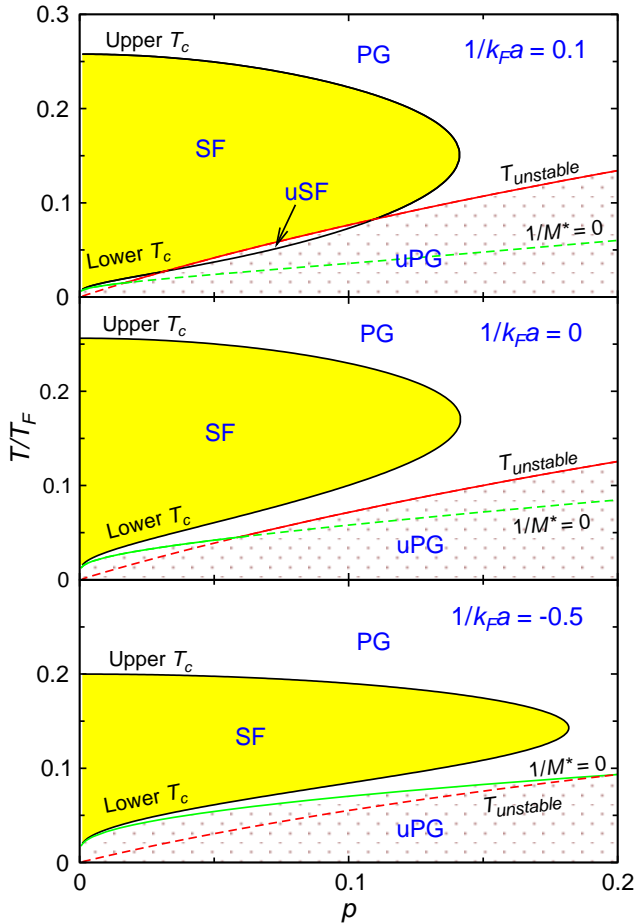}
\caption{Phase diagram on the $T$-$p$ plane on the BCS side of the
  $T_c=0$ line in Fig.~\ref{fig:PhaseT0}, for $1/k_Fa=0.1$, 0, and -0.5,
  from top to bottom. In each panel, the superfluid region is shaded in
  yellow, on the left side of the $T_c$ curve.  Except for $1/k_Fa=0.1$,
  the (red) $T_{unstable}$ line does not intersect with the $T_c$
  curve. There is a small unstable superfluid phase (uSF) for
  $1/k_Fa=0.1$, as labeled in the top panel. For $1/k_Fa=-0.5$,
  the pseudogap (PG) is present only in a narrow temperature range since
  here the pair formation temperature $T^*$ (approximately given by
  $T_c^{MF}$) is not much higher than the upper $T_c$. At higher $T$,
  the phase is a polarized normal Fermi gas.  Slightly below the lower
  $T_c$, the pair mass becomes negative (below the green $1/M^*=0$
  line), implying that the paired phase in the mean-field treatment is
  unstable in this region. The $M^*>0$ requirement is a tighter
  constraint than Eq.~(\ref{eq:d2D}) in the BCS regime at low $T$ and
  low $p$. }
\label{fig:LowerTc}
\end{figure}

The next two figures correspond to regions on either side of the quantum
critical point $T_c =0$ in Figs.~\ref{fig:PhaseT0} and
~\ref{fig:PhaseT1}.  The region to the right of this line (``Regions IIB
and IID" in Ref.~\cite{Chien06}) is further discussed in
Fig.~\ref{fig:T_unstable}. It corresponds to the BEC side of resonance
above $1/k_F a = 0.2$.  On the left of this line (``Regions IIA and IIC"
in Ref.~\cite{Chien06}) which includes the BCS side of resonance as well
as the unitary phase we summarize our findings in Fig.~\ref{fig:LowerTc}
below.

In Fig.~\ref{fig:T_unstable}, we show how $T_c$ and $T_{unstable}$
evolve with $p$ and $1/k_Fa$ on the BEC side, where $T_{unstable}$ is
the temperature where $\frac{\p^2 \Omega}{\p\Delta^2}=0$. We do not show
the deep BEC case where the superfluid phase is always stable.
Here we plot these two families of curves 
for $1/k_Fa=0.2$, 0.3, and 1.0, from top to bottom.  The upper and lower
curves in each panel are for $T_c$ and for $T_{unstable}$, respectively.
Each phase diagram is composed of four different phases. Inside the
shaded area, above $T_{unstable}$ but below $T_c$, there is a stable
intermediate temperature superfluid.  Below these two curves, the
superfluid phase is unstable and this is denoted by ``uSF''. To the
right of the $T_c$ curve we have a stable pseudogap phase above the
$T_{unstable}$ line, labeled by ``PG''.  Below the $T_{unstable}$ line
we have an unstable pseudogap phase labeled ``uPG''.  When there is a PG
phase (stable or unstable) the pairing onset temperature $T^*$
(approximately given by $T_c^{MF}$, not shown) is higher than $ T_c$.
As $1/k_Fa$ increases, the intersection point between $T_{unstable}$ and
the $T=0$ axis moves to the right, and the area of the stable superfluid
phase grows, until $1/k_Fa \approx 2.3$, where the area of the unstable
phase shrinks to zero. The case $1/k_Fa=0.2$ is unusual; the $T_c$ curve
bends to the left at low temperature. We also show the $n_s=0$ line for
$1/k_Fa=0.2$. It is completely within the unstable regime, and it
disappears for $1/k_Fa \gtrsim 0.5$. The pair mass is always positive
for $1/k_Fa>0.2$.

It should be noted that at a given $T$ below the point where $T_c
=T_{unstable}$, there is a critical polarization $p_c$, above which the
superfluid becomes unstable against phase separation (or an FFLO-like
pairing state). From Fig.~\ref{fig:T_unstable}, it is evident that $p_c$
increases with $T$.  This trend is consistent with recent experimental
results reported by the Rice group \cite{Hulet_PC}, which report that
the critical polarization for phase separation decreases with decreasing
$T$. In the present work, of course, we have not incorporated the
effects of the trap potential.  These were discussed elsewhere
\cite{ChienRapid}.

We now address Fig.~\ref{fig:LowerTc} which presents similar plots in
the $T$-$p$ plane for fixed values of $1/k_Fa = 0.1,0$, and -0.5.  This
corresponds to the left side of the $T_c=0$ curve in
Fig.~\ref{fig:PhaseT0}.  Generally, the $T_{unstable}$ curve lies below
the corresponding lower $T_c$ curve. The shaded region on the left of
the $T_c$ curve in each of the three panels is a stable superfluid
phase. For $1/k_Fa=0.1$, however, the superfluid phase has a tiny region
of instability enclosed by the $T_c$ and $T_{unstable}$ curves. As in
Fig.~\ref{fig:T_unstable}, the normal region is split into a stable and
unstable phase, where pseudogap effects persist except at high $T$. In
all cases, the paired phase below $T_{unstable}$ is unstable. The
effective pair mass becomes negative below the line defined by
$1/M^*=0$, which intersects the $T_{unstable}$ line. Here the positivity
of the pair mass provides a more stringent constraint on stability than
Eq.~(\ref{eq:d2D}).  Finally, just as in Fig.~\ref{fig:T_unstable}, one
can find a critical polarization $p_c$ for these values of $1/k_Fa$ as
well, and $p_c$ increases with $T$.

It should be noted that Figs. \ref{fig:T_unstable} and \ref{fig:LowerTc}
evolve smoothly into each other.  As $1/k_Fa$ changes continuously from
0.2 to 0.1, the $T_c$ curve at low $T$ in the top panel in
Figs. \ref{fig:T_unstable} will bend further to the left, and eventually
touch the $p=0$ axis and change into the top panel of
Fig.~\ref{fig:LowerTc}.

\begin{figure}
\includegraphics[clip]{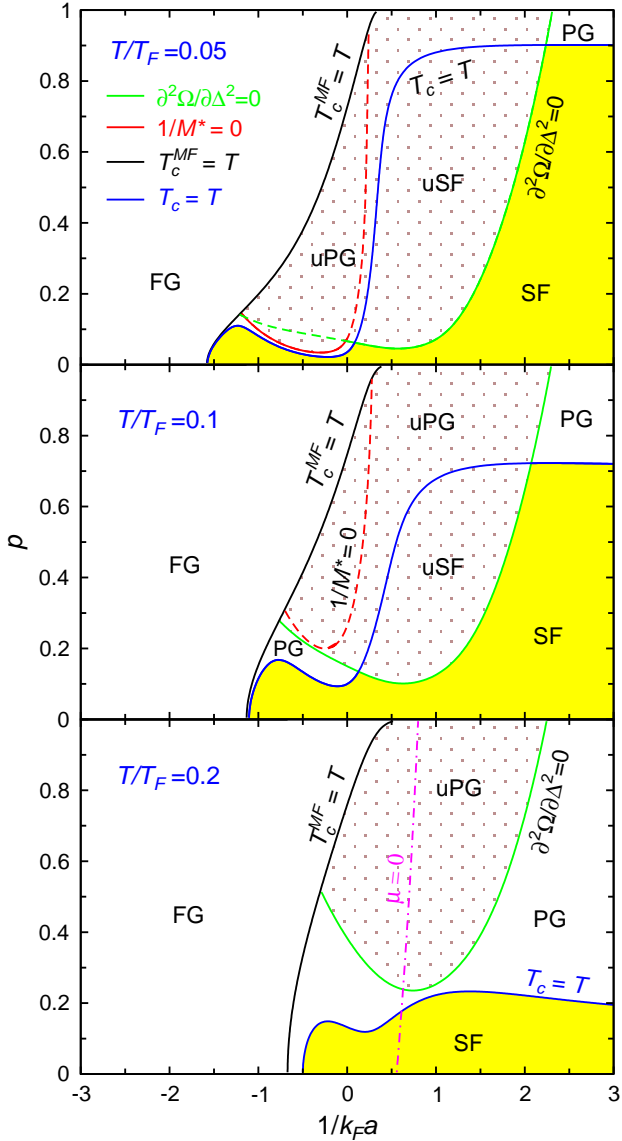}
\caption{Evolution of the phase diagram in the $p-1/k_Fa$ plane with
  temperature. From top to bottom, $T/T_F=0.05$, 0.1, and 0.2,
  respectively. The phase diagrams are split into different phases by
  the solid curves, as labeled in the figure. Except the $\mu=0$ line
  (shown for $T/T_F=0.2$ only) stay relatively unchanged, all other
  phase boundaries moves with $T$. The mean-field ``phase transition''
  line, defined by $T_c^{MF}=T$, moves to the right. The (blue) $T_c=T$
  line separates the superfluid and the pseudogap (or normal) phase, and
  the (green) $\p^2\Omega/\p\Delta^2=0$ line, in conjunction with the
  (red) $1/M^*=0$ line, separates the stable and unstable phases. Stable
  superfluid exists in the yellow shaded region, the dotted region is
  either an unstable superfluid (labeled by ``uSF'') or unstable
  pseudogap (``uPG'') phase. The white space on the right of
  $T_c^{MF}=T$ line is a stable pseudogap (``PG'') phase. The $1/M^*>0$
  represents a stronger constraint than Eq.~(\ref{eq:d2D}) only at low
  $T$ and low $p$ (e.g., in the top panel). The $1/M^*=0$ line moves to
  the upper right and disappears at slightly below $T = 0.2$.  The sign
  change of $n_s^{MF}$ (not shown) always occurs within the unstable
  regions. We also show $\mu=0$ which separates fermionic from bosonic
  regimes.  }
\label{fig:PhaseT}
\end{figure}

We end with Fig.~\ref{fig:PhaseT}. This shows how the various phase
boundaries first presented in Fig.~\ref{fig:PhaseT0} evolve with
temperature. The three panels, from top to bottom, correspond to three
different temperatures $T/T_F=0.05$, 0.1, and 0.2, respectively. The
\emph{solid lines} split the phase diagram into a few distinct phases.
The points at which $T_c^{MF}$ is equal to the given temperature of each
figure lead to a line which separates the unpaired Fermi gas (FG) phase
from the paired phases. Similarly, the $T_c=T$ line separates the
superfluid phase (below the line) from the pseudogap or normal phases
(above the line). Finally, the lines associated with
$\p^2\Omega/\p\Delta^2=0$ and $1/M^*=0$ separate the stable superfluid
and pseudogap phases (below this line) from unstable phases. These two
lines intersect with each other, and when this happens, one will appear
partly inside the unstable region defined by the other, as indicated by
the red and green dashed lines. Thus the stable superfluid phase lies in
the yellow shaded region. The $n_s^{MF}=0$ line (not shown) always
appears completely within the unstable phases, and therefore, does not
provide a separate phase boundary, as in the $T=0$ case. (Note here we
use $n_s^{MF}$ instead of $n_s$ because $n_s=0$ outside the superfluid
regions.)

As temperature increases, the line associated with $T_c^{MF}=T$ moves to
the right. At the same time both the $1/M^*=0$ curve and the
$n_s^{MF}=0$ curve shrink and move to the upper left, ultimately
disappearing slightly below $T/T_F=0.2$. The $\mu=0$ boundary between
fermionic and bosonic regimes is rather insensitive to $T$, and so is
indicated just for the case of the highest temperature.

The evolution of the $T_c=T$ and $\p^2\Omega/\p\Delta^2=0$ curves is
rather interesting. For low polarization $p$, both curves move to the
left into the BCS regime. This reflects the fact that temperature
stabilizes the paired and superfluid phase. This is consistent with our
observation of intermediate temperature superfluidity in the presence of
population imbalance. The $T_c = T$ curve on the BEC side is
continuously suppressed by raising $T$.  This is simply due to fact that
$T_c$ decreases with $p$ for given $1/k_Fa$ in this regime.  In contrast
with the $T=0$ case, we have a stable pseudogap phase appearing at
finite $T$. This corresponds to the white region on the right of the
$T_c^{MF}=T$ line, below the $\p^2\Omega/\p\Delta^2=0$ and $1/M^*=0$
lines, and above the $T_c=T$ line. At $T/T_F=0.2$, the
$\p^2\Omega/\p\Delta^2=0$ and the $T_c=T$ lines do not intersect each
other, so that the entire superfluid phase is stable.

\section{Conclusion}

In summary, in this paper we have studied the stability conditions for
polarized fermionic superfluids in considerable detail.  We have
restricted our attention to zero momentum condensate pairs, thereby,
excluding FFLO-like phases.  We find that the positive definiteness of
the number susceptibility matrix is equivalent to the positivity of
$(\p^2\Omega/\p\Delta^2)_{\mu,h}$, provided the gap equation
(\ref{eq:gap}) is satisfied. In addition, we find that the positivity
requirement of the effective pair mass constitutes another nontrivial
stability condition. At low temperature and low population imbalance in
the fermionic regime, this latter condition may be more stringent that
the positivity requirement of $(\p^2\Omega/\p\Delta^2)_{\mu,h}$. 

To present our results in a more concrete fashion, we study the phase
diagram in different planes of the three-dimensional space spanned by
the parameters ($T$, $p$, $1/k_Fa$). In particular, we have shown how
the phase boundaries defined by the various stability conditions evolve
with temperature.  At relatively low polarization $p\lesssim 0.2$, there
exists a stable superfluid phase at finite $T$, which may not be stable
at $T=0$. In a related fashion, we have shown where on the phase diagram
pairing without condensation (that is, a pseudogap phase) appears.

A general observation associated with these population imbalanced
theories (which exclude the FFLO phase) is that in the important regime
near unitarity, superfluidity only appears at intermediate temperatures
\cite{Chien06}.  This was seen earlier in mean field approaches
\cite{Sedrakian} via studies of the quantity we define as $T_c^{MF}$,
which was found to be double valued.  In the present framework (which
goes beyond by including pairing fluctuation effects), this intermediate
temperature superfluidity appears via the presence of two $T_c's$ at
unitarity.  Somewhat above unitarity, and towards the BEC regime, we
find another mechanism for arriving at intermediate temperature
superfluidity. This can occur due to a low temperature instability
towards phase separation, associated with a negative sign in
$(\p^2\Omega/\p\Delta^2)_{\mu,h}$.

While $(\p^2\Omega/\p\Delta^2)_{\mu,h} < 0 $ is argued to lead to phase
separation, one may imagine that a negative pair mass would result in a
lower pair energy at a finite momentum than at $\textbf{q}_0 = 0$ and
thus a possible LOFF-like condensate. On the other hand, a negative
$n_s$ has also been argued \cite{Tsinghua} to be associated with a LOFF
phase. Future studies are needed to unravel these various possible
phases within the \emph{unstable} regimes in the phase diagrams.

\acknowledgments

This work was supported by NSF Grant No.~PHY-0555325 and NSF-MRSEC Grant
No.~DMR-0213745

\bibliographystyle{apsrev}


\end{document}